\newif\ifmnras
\mnrasfalse
\ifmnras
	\documentclass[useAMS,usenatbib]{mn2e}
\else
	\documentclass[apj,numberedappendix]{emulateapj}
\fi
\usepackage{times}
\usepackage{graphics,epsf}
\usepackage{amsmath}                
\usepackage{amsfonts}               
\usepackage{amssymb}                
\usepackage{epsfig}                 
\usepackage{graphicx}
\usepackage{url}
\usepackage{tabularx}
\usepackage{booktabs}
\usepackage{hyperref}

\def \cm{~\mathrm{cm}}
\def \s{~\mathrm{s}}

\def \km{~\mathrm{km}}

\def \cms{~\mathrm{cm}~\mathrm{s}^{-1}}

\def \K{~\mathrm{K}}

\def \erg{~\mathrm{erg}}

\def \yr{~\mathrm{yr}}
\def \myr{~\mathrm{Myr}}

\def \zams{\mathrm{ZAMS}}

\ifmnras
	\def \aap{A\&A}
	\def \apj{ApJ}
	\def \apjl{ApJ}
	\def \apjs{ApJS}
	\def \nat{Nature}
	\def \mnras{MNRAS}
\fi

\def \raa{RAA}

\ifmnras
	\title[Angular momentum in convective He shell]{Angular momentum fluctuations in the convective helium shell of massive stars}

	\author[A. Gilkis \& N. Soker]{Avishai Gilkis and Noam Soker \\
	Department of Physics, Technion -- Israel, Institute of Technology, Haifa 32000, Israel;
	agilkis@tx.technion.ac.il;
	soker@physics.technion.ac.il}
\fi

\begin{document}

\ifmnras
	\pagerange{\pageref{firstpage}--\pageref{lastpage}} \pubyear{2015}

	\maketitle
\else
	\title{Angular momentum fluctuations in the convective helium shell of massive stars}

	\author{Avishai Gilkis}
	\author{Noam Soker}
	\affil{Department of Physics, Technion -- Israel
	Institute of Technology, Haifa 32000, Israel;
	agilkis@tx.technion.ac.il; soker@physics.technion.ac.il}
\fi

\label{firstpage}

\begin{abstract}
We find significant fluctuations of angular momentum 
within the convective helium shell of a pre-collapse massive star --
a core-collapse supernova progenitor --
which may facilitate the formation of accretion disks and jets that can explode the star.
The convective flow in our model of an evolved $M_\mathrm{ZAMS}=15M_{\odot}$ star,
computed with the sub-sonic hydrodynamic solver MAESTRO,
contains entire shells with net angular momentum in different directions.
This phenomenon may have important implications for the late evolutionary stages of massive stars,
and for the dynamics of core-collapse.

\end{abstract}

\section{INTRODUCTION}
\label{sec:introduction}

Theory suggests that massive enough stars develop a core of iron-group elements,
which upon reaching a critical mass undergoes a catastrophic collapse followed by
an energetic and luminous explosion termed a core-collapse supernova (CCSN).
Although stars are close to being spherical objects,
there are phenomena for which a spherically symmetric one-dimensional treatment may be insufficient,
both in the nearly hydrostatic states during the long stellar lifetime prior to collapse,
and in the hydrodynamic explosion itself.

In stellar evolution modeling,
convective mixing and transfer of energy is a notably multi-dimensional phenomenon
(e.g., \citealt{Beeck2015}).
Convection is usually treated using the approximate semi-empirical Mixing-Length Theory (MLT).
This simplified treatment might be inadequate in some cases.
Ongoing studies aim at improving the theoretical modeling and understanding of convection in stellar interiors
(e.g., \citealt{Meakin2007,Arnett2009,Arnett2015,Viallet2013,Viallet2015,Viallet2016}).

Examining explosion models,
the long-studied neutrino-driven mechanism \citep{Colgate1966,bethe1985}
fails in most one-dimensional simulations
(e.g., \citealt{Burrows1995,Rampp2000,Mezzacappa2001,Liebend2005}),
and it might be important to look at multi-dimensional effects such as
the standing accretion-shock instability
(SASI; e.g., \citealt{BlondinMezzacappa2003, BlondinMezzacappa2007, Fernandez2010}).
\cite{PapishNordhausSoker2015} suggest that neutrino-driven mechanisms might not work.

The multi-dimensional phenomena in evolution and explosion modeling may be interconnected.
Recent studies \citep{CouchOtt2013,CouchOtt2015,MuellerJanka2015}
found that non-spherical perturbations in the progenitor may increase post-collapse turbulent pressure,
significantly influencing the dynamic behavior in the shocked region surrounding the forming NS.
In these studies,
large velocity fluctuations were introduced in the pre-collapse core.
In an earlier paper \citep{GilkisSoker2015}
we found that such velocity perturbations imply sufficiently large specific angular momentum fluctuations
for the formation of intermittent accretion disks and a jittering-jets explosion.

Among the alternatives to the neutrino-driven explosion mechanism
(see \citealt{Janka2012} for a review),
most prominent are jet-driven explosions (e.g., \citealt{LeBlanc1970,Khokhlov1999,Lazzati2012}).
Another interesting mechanism is the collapse-induced thermonuclear explosion \citep{KushnirKatz2014,Kushnir2015a,Kushnir2015b}.
The jet-driven scenario is inherently multi-dimensional, where a pre-collapse rapid core rotation may facilitate
the formation of magnetorotationally driven bipolar outflows \citep{Mosta2015}.
We note that the above cited works require a rapidly rotating pre-collapse core,
and hence limited to a small fraction of all CCSNe.

A different approach of jet-driven explosions
tries to explain all energetic CCSNe, for cases of rapid as well as slow stellar rotation
\citep{Soker2010,PapishSoker2011,PapishSoker2012a,PapishSoker2012b,PapishSoker2014a,PapishSoker2014b}.
In this model
accretion from material with varying angular momentum in a collapsing star
leads to the formation of accretion disks with varying axis direction.
These intermittent accretion disks launch jets with different directions,
termed `jittering jets',
that might explode the star with the observed energy.
The jittering jets model faces the challenge of supplying the required angular momentum for an accretion disk,
and consequently forming jets.
The required angular momentum for jet formation may be lower than usually implied,
when a dynamo operates in a shear layer around the proto-neutron star \citep{Akiyama2003,Schreier2016}.


Recent studies have concentrated on
non-spherical perturbations in the close vicinity of the iron core just prior to collapse
(e.g., \citealt{Couchetal2015}).
We look a bit earlier in the stellar evolution
and further out in the star for a possible source of angular momentum.
While the outer hydrogen region probably has tremendous fluctuations of angular momentum,
the sudden change in the gravitational
potential as a result of neutrino emission during NS formation, results
in the expulsion of the envelope \citep{Lovegrove2013}.
Analytic estimations show that the convective helium shell of an evolved massive star may
have large fluctuations of angular momentum as well \citep{GilkisSoker2014}.
In a scenario where after core-collapse the inner silicon/oxygen regions
fail to produce a sufficient driving force for a supernova explosion,
accretion from the helium region might result in the formation of intermittent accretion disks and jets.
Accretion from material with varying angular momentum in a collapsing star can lead to an energetic supernova explosion
in the jittering jets model.

In the present study we explore the properties of the pre-collapse convective helium region.
We perform a three-dimensional hydrodynamic simulation of the convective flow,
strengthening our preliminary analytic estimations.
The helium region does not change much in the relatively short time left until collapse,
so that our results are relevant for the final stage of the stellar evolution.
In section \ref{sec:setup} we describe the numerical setup and method.
Our main results are presented in section \ref{sec:results}.
Some implications for core-collapse are concisely discussed in section \ref{sec:collapse}.
We summarize in section \ref{sec:summary}.


\section{NUMERICAL SETUP}
\label{sec:setup}

\begin{figure}
   \centering
    \includegraphics*[scale=0.54]{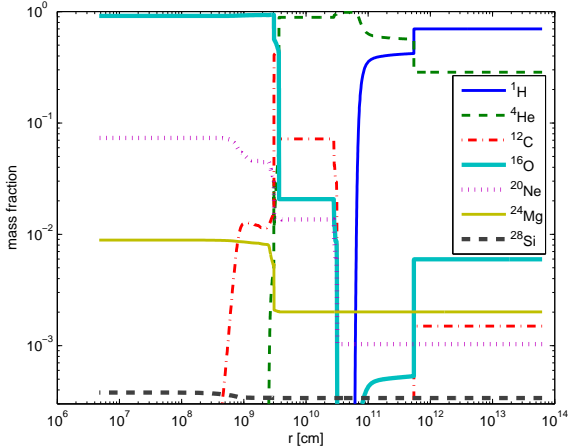} \\
      \caption{Detailed composition for our stellar model
      of a non-rotating $M_\zams=15 M_\odot$ star with metallicity of $Z=0.014$.
      The profile is presented at an age of $t= 12.5 \myr$, about $5.5 \yr$ before explosion.
      The stellar parameters at this stage are as follows.
      Stellar mass of $M=13.3 M_\odot$,
      effective temperature of $T=3235K$,
      photospheric radius of $R=902R_\odot$ and luminosity of $L=8 \times 10^4 L_\odot$.
      The convective helium shell simulated here extends from $r_\mathrm{in} = 3.7 \times 10^9 \cm $ to $r_\mathrm{out} = 2.8 \times 10^{10} \cm$.
      Results below will all be presented at this time and for this model.}
      \label{fig:composition}
\end{figure}

We use a stellar model constructed by 
Modules for Experiments in Stellar Astrophysics (MESA version 5819; \citealt{Paxton2011,Paxton2013}),
with an initial mass of $M_\zams=15 M_\odot$ and metallicity of $Z=0.014$.
Due to stellar winds calculated here with the so-called `Dutch' scheme (e.g., \citealt{Nugis2000,Vink2001})
the final mass is $13.3M_\odot$.
Effects of rotation and magnetic fields are neglected.
We examine the star during the helium shell burning stage,
about 5.5 years
prior to collapse.
At this stage, the star is a red supergiant (RSG)
with an effective surface temperature of $T=3235K$,
photospheric radius of $R=902R_\odot$
and luminosity of $L=8 \times 10^4 L_\odot$.
The detailed composition of the stellar model is shown in Figure \ref{fig:composition}.

There are two reasons for simulating the star at this time.
($i$)
The convective helium shell may be of interest as a potential supply of material with angular momentum for accretion
onto the newly-formed compact object following core-collapse,
in a scenario where the inner shells fail in driving a successful explosion.
The convective hydrogen envelope is less relevant,
as it is likely to be expelled due to decrease of the NS gravitational mass 
by neutrino cooling \citep{Nadezhin1980,Lovegrove2013}.
The structure outside $r_\mathrm{O-core}\approx 3.5 \times 10^{9} \cm$
remains essentially unchanged until core-collapse,
so the results presented here are relevant for the core-collapse epoch.
If we had taken an earlier point of the stellar evolution,
the different core structure would have lessened the pertinence of our simulation.
($ii$)
While in the final stage before core-collapse there are several convective shells
(silicon and oxygen, in addition to helium and hydrogen at the simulated stage),
the complex structure would have made the simulation more difficult to run.
We hope the methods detailed here will inspire future works on
angular momentum fluctuations in the silicon and oxygen shells
in core-collapse supernova progenitors.

Convection in MESA is according to the Mixing-Length Theory (MLT),
which gives for our model a convective outer hydrogen envelope,
and a convective helium region surrounding a radiative oxygen core.
The outer part of the helium region is radiative as well.
In this study we are not interested in the outer envelope,
and focus on the convective helium region,
extending from $r_\mathrm{in} = 3.7 \times 10^9 \cm $  and $M_\mathrm{in} = 2.4 M_\odot$, 
to $r_\mathrm{out} = 2.8 \times 10^{10} \cm$ and $M_\mathrm{out} =3.7 M_\odot$.
If and when this region collapses to the core,
the free fall time from the inner and outer boundaries of the helium convective regions are $t_\mathrm{ff-in} =14 \s$,
and  $t_\mathrm{ff-out} =228 \s$, respectively.
Interestingly, this range is compatible with the duration of long gamma ray bursts (LGRB).
This is not the case studied here,
but the behavior of the helium convective layer might be relevant to the formation of jets in GRBs.

The inner region of the one-dimensional MESA model is mapped into a three-dimensional
grid in the low Mach number stellar hydrodynamics code MAESTRO \citep{Nonaka2010}.
The MAESTRO code was developed mainly for the purpose of studying
convection in white dwarfs experiencing thermonuclear deflagration
\citep{Almgren2006a,Almgren2006b,Almgren2008, Zingale2009},
though more recently also for modeling convection preceding Type I X-ray Bursts \citep{Malone2011,Malone2014,Zingale2015}
and convection in massive stars \citep{Gilet2013}.
This latter study motivated us to use MAESTRO for our purpose --
the study of angular momentum fluctuations in a convective region of a massive star.
For this purpose we need a full sphere domain,
with no imposed reflective boundaries for reduction of computational costs.

Similarly to \cite{Gilet2013},
we initialize the velocity field with small perturbations
(much lower in magnitude than the expected, and obtained, flow velocities),
and use velocity damping near the edges of the simulation domain (far from the regions of interest).
We further added velocity damping in the oxygen core,
as we encountered numerical problems in this region of the simulation -- spurious velocities and heating.
This may be due to the absence of radiative transfer in the simulation (important in the non-convective core)
or because of insufficient resolution.
We ran simulations with damping of the entire core, part of it, and no damping at all.
The simulated convective helium region changed little between the simulations,
and the inner oxygen core is of no particular interest in the present study.
We therefore conclude that our results are robust,
although we may need to address this issue in a future work.

We have incorporated an approximate reaction network in the code,
which includes only the triple-alpha process
(the only relevant process in the region of interest).
The energy generation rate is according to \cite*{Kippenhahn2012}, p.~197,
\begin{multline}
\epsilon_{3\alpha} = 6.272 \rho^2 X_4^3 \cdot \left( 1 + 0.00158 T_9^{-0.65} \right) \\
\times [ 2.43 \times 10^9 T_9^{-2/3} \exp \left( -13.490 T_9^{-1/3} - \left( T_9 / 0.15 \right)^2 \right) \\
\times \left( 1 + 74.5 T_9 \right)
+ 6.09 \times 10^5 T_9^{-3/2} \exp \left( -1.054 / T_9 \right) ] \\
\times [ 2.76 \times 10^7 T_9^{-2/3} \exp \left( -23.570 T_9^{-1/3} - \left( T_9 / 0.4 \right)^2 \right) \\
\times \left( 1 + 5.47 T_9 + 326 T_9^2 \right) + 130.7 T_9^{-3/2} \exp \left( -3.338 / T_9 \right) \\
+ 2.51 \times 10^4 T_9^{-3/2} \exp \left( -20.307 / T_9 \right) ] ,
\label{eq:triplealpha}
\end{multline}
where $\epsilon_{3\alpha}$ is in $ \mathrm{erg}~\mathrm{gr}^{-1}~\s^{-1}$,
$\rho$ is the density, $X_4$ is the helium mass fraction,
and $T_9$ is the temperature divided by $10^9 \K$.
We have neglected composition changes,
as the nuclear timescale is significantly longer than the dynamical timescale.

The Adaptive Mesh Refinement (AMR) capabilities of MAESTRO
were employed to keep a grid resolution better than $0.1 H_P$
throughout the convective helium region, where $H_P$ is
the pressure scale height.
We performed several simulations with lower finest resolution
(higher resolutions are prohibited with our current computational resources).
The magnitude of the angular momentum, which was our primary concern,
is similar in the different simulations.
The maximal Mach number of the convective motion and the average velocity,
however, increase with decreasing resolution.
Clearly, simulations with higher resolution are required to fully understand the phenomena investigated here,
but for this introductory study our results are sufficient.
A detailed analysis of the lower-resolution full simulations is presented in Appendix \ref{app:low},
and additional simulations of one eighth of the domain and higher resolution are presented in Appendix \ref{app:oct}.

\section{SIMULATION RESULTS}
\label{sec:results}

\subsection{Angular momentum}
\label{subsec:j}

\begin{figure}
   \centering
    \includegraphics*[scale=0.23]{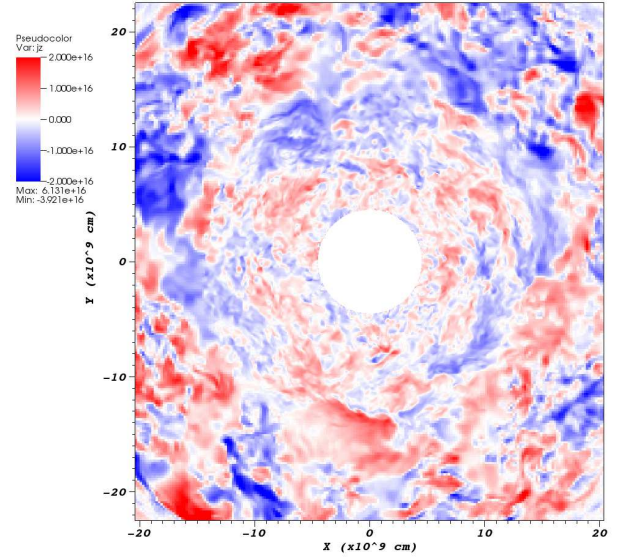} \\
      \caption{ Specific angular momentum
      in the $xy$-plane at $t=20000 \s$ in the simulation.
      Blue-shaded regions have a clockwise azimuthal velocity component ($j_z<0$), 
      while red-shaded regions have a counterclockwise azimuthal velocity component ($j_z>0$).
      White regions have no azimuthal velocity component ($j_z=0$).
      The color-coding runs from $ {-2} \times10^{16} \cm^2 \s^{-1}$ up to $2\times10^{16} \cm^2 \s^{-1}$,
      although the specific angular momentum exceeds this value at some points.
      The inner oxygen core is omitted from the figure.}
      \label{fig:velocitymap}
\end{figure}

\begin{figure}
   \centering
   \vspace*{0.1 cm}
    \includegraphics*[scale=0.23]{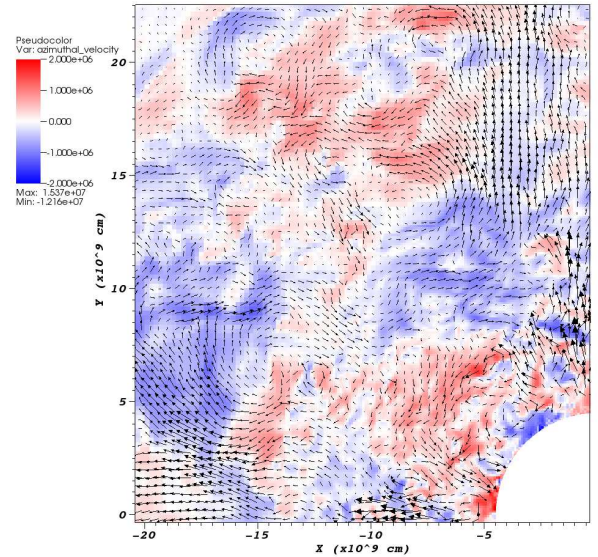} \\
      \caption{Azimuthal velocity color-map for a close-up of a region from Figure \ref{fig:velocitymap},
      with added velocity arrows for illustration of the vortices in the flow.
      The color-coding runs from an azimuthal velocity of $ {-2} \times10^6 \cm \s^{-1}$ up to  $2\times10^6 \cm \s^{-1}$.
	  Similarly to Figure \ref{fig:velocitymap},
	  blue-shaded and red-shaded regions have clockwise and counterclockwise velocity components, respectively.
      The velocity arrows are scaled -- larger arrows represent higher velocities
      (an arrow spanning $10^9 \cm$ on the plot, for example, represents a velocity of $2\times 10^6 \cms$).}
      \label{fig:velocitycloseup}
\end{figure}

We ran the simulation for a physical time of close to  $t=20000 \s$.
This is about 85 times the free fall time from the outer convective boundary $t_\mathrm{ff-out}$,
and about 8 times the typical convective overturn time $t_\mathrm{conv} \equiv D_\mathrm{conv}/v_\mathrm{conv} \approx 2500 \s $.
The latter is calculated from the the typical diameter of the vortices $D_\mathrm{conv} \approx 5 \times 10^9 \cm$,
and the typical velocity in the vortices of $v_\mathrm{conv} \approx 2\times 10^6 \cm \s^{-1}$,
as seen in figures \ref{fig:velocitymap} and \ref{fig:velocitycloseup} presented at the end of the simulation.
Figure \ref{fig:velocitymap} shows the flow morphology,
with color indicating the local specific angular momentum calculated from the flow velocity.
Figure \ref{fig:velocitycloseup} shows a close-up of a region from Figure \ref{fig:velocitymap},
with color indicating the magnitude and sense of the velocity,
to emphasize the large-scale vortices that develop in the core.

\begin{figure}
   \centering
    \includegraphics*[scale=0.54]{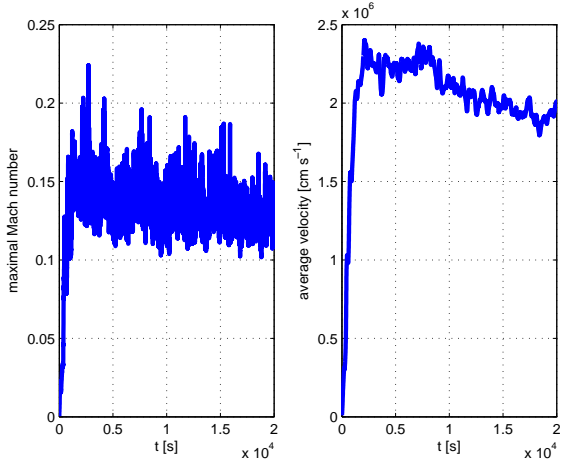} \\
      \caption{\textit{Left:} Maximal Mach number in the simulation domain.
      The values reached in the simulation, of $\mathcal{M} \approx 0.15$,
      are two orders of magnitude above the mixing-length theory
      value for this model ($\mathcal{M}_\mathrm{MLT} \approx 10^{-3}$).
      \textit{Right:} Average velocity in the simulation domain.
      This too is much higher than the velocities of the MLT,
      which are around $v_\mathrm{MLT} \approx 10^5-2\times 10^5 \cms$
      for our stellar model.
      The disparity to the MLT is smaller for the averaged velocity,
      as there is an extensive range of velocities in the flow field,
      and the maximal velocities (and Mach numbers) reach values significantly higher than the mean.}
      \label{fig:machv}
\end{figure}

Figure \ref{fig:machv} shows the evolution with time of 
the global computational diagnostics of the maximal Mach number and the averaged velocity magnitude.
It can be seen that after leaving the initial close-to-zero velocity setup,
a steady flow develops after a time of about $t_\mathrm{conv}$. 
Another important conclusion drawn from this figure is that
the typical convective velocities are much higher than values given by the mixing-length theory.
Similar results of higher than the MLT velocities have been obtained by, e.g., \cite{Arnett2009}.
This result holds even when a velocity lower by a factor of 2 is taken,
as might be the case with higher resolution (see Appendix \ref{app:oct}).

\begin{figure*}
   \centering
    \includegraphics*[scale=0.34]{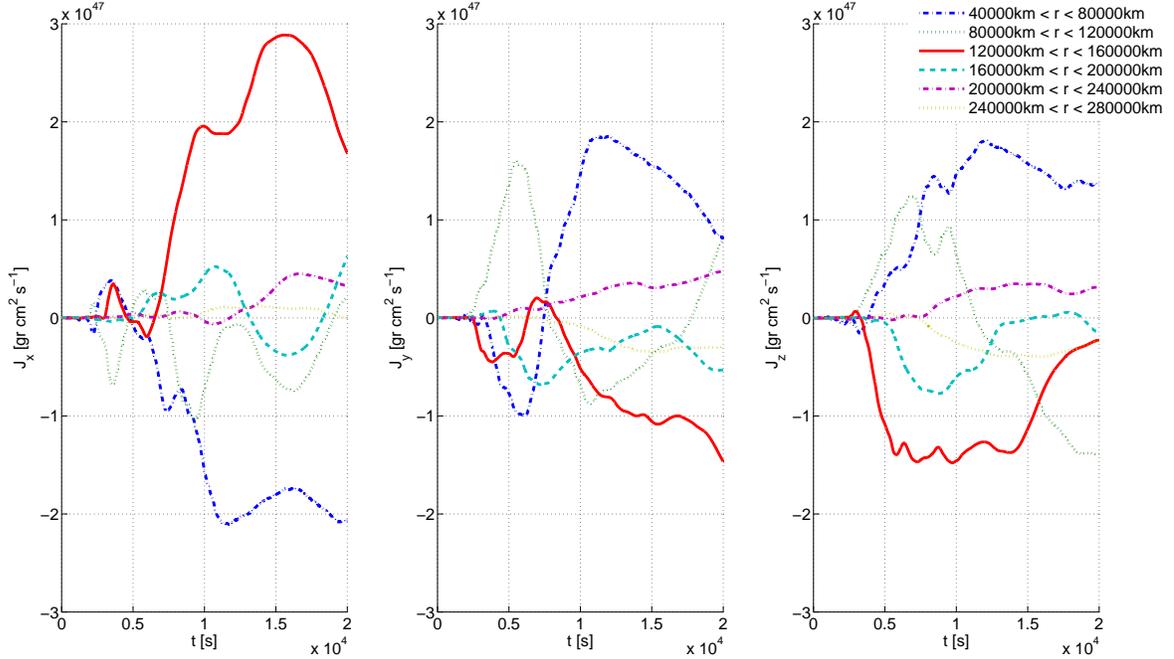} \\
      \caption{Angular momentum components of several shells within the convective region throughout the simulation time.
      The location of the shells is indicated in the inset in the right panel. 
      Several shells possess net angular momentum with large values and in different directions.
      The masses of the shells from the inner shell to the outer are
      $0.25$, $0.25$, $0.23$, $0.22$, $0.19$ and $0.17 M_\odot$.
      The core mass inner to the helium shell is $M_\mathrm{in} =2.4 M_\odot$.
      The average specific angular momentum is therefore approximately $\mathrm{few} \times 10^{14} \cm^2 \s^{-1}$,
      which is lower than the specific angular momentum for a Keplerian orbit around a NS or BH.
      However, the partition into shells with widths of $\Delta r=40000 \km$ is arbitrary.
      While we see here the large-scale behavior,
      our simulation might not capture some finer details relevant for post-collapse dynamics, as
      the differences in the free-fall times between the shell boundaries correspond
      to thousands of orbital times around the newly-formed NS or BH.
      A more focused analysis of the angular momentum distribution within shells
      at one time is shown in Figure \ref{fig:jspecific}.}
      \label{fig:jfluctutations}
\end{figure*}

\begin{figure}
   \centering
    \includegraphics*[scale=0.54]{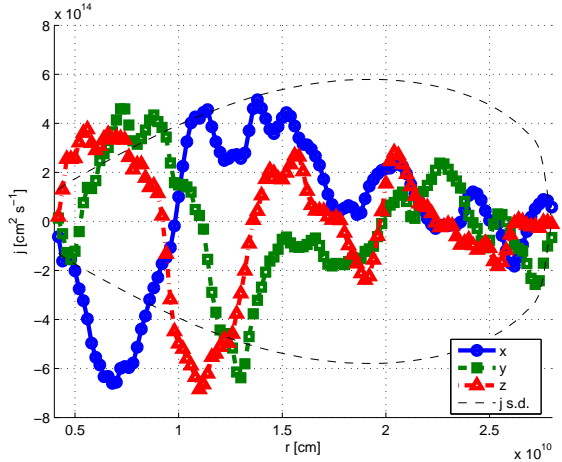} \\
      \caption{Specific angular momentum components in thins shells of $\Delta r=2000 \km$ at  $t=20000 \s$
      in the convective zone in the simulation, between $r=40000\km$ and $r=280000\km$.
      Black dashed lines show an analytic approximation
      of the standard deviation of specific angular momentum in the MLT,
      following \cite{GilkisSoker2014}
      so that
      $\sigma\left(j\right) = \left(2/27\right)^{1/2}\left(\Delta r\right)^{-1/2} a_c^{3/2} v_c$,
      with $a_c$ taken as half the mixing-length and $v_c$ the MLT convection velocity.
      The specific angular momentum for a Keplerian orbit around a newly-formed NS or BH
      is approximately $j_\mathrm{Kep} \approx \mathrm{few} \times 10^{16} \cm^2 \s^{-1}$;
      the values for the thin shells shown here is lower, reaching
      $j_\mathrm{max}\left(\mathrm{fluctuations}\right) \approx 6.85 \times 10^{14} \cm^2 \s^{-1} \approx 0.03 j_\mathrm{Kep}$.
      Although the specific angular momentum is lower than the Keplerian,
      it is not negligible, and a belt-like structure might form around the NS or BH
      (as discussed by \citealt{PapishGilkisSoker2015}).
      The difference in free-fall times for $\Delta r=2000 \km$ is $1-3 \s$,
      corresponding to hundreds of orbital times around the newly-formed compact object,
      so there is sufficient time to form accretion belts and jets upon core-collapse.}
      \label{fig:jspecific}
\end{figure}

Figure \ref{fig:jfluctutations} shows the evolution of the angular momentum in several shells.
We notice the significant net angular momentum that is developed in the shells,
and the varying direction of the angular momentum axis.
In Figure \ref{fig:jspecific} we show the specific angular momentum profile of the material within the inner part of the convective region
at $t=20000 \s$ in the simulation. 
In plotting Figure \ref{fig:jspecific} we used shells of width $\Delta r=2000 \km$,
much thinner than the $\Delta r = 40000 \km$ shells used in drawing Figure \ref{fig:jfluctutations}.
The difference in free fall times between the two boundaries of the thin shells is $t_{ff} \approx 1-3 \s$,
equivalent to thousands of dynamical times on the surface of the newly formed NS (or BH).
Shorter time scales may be of interest for the post-collapse dynamics,
but a higher resolution is required, accordingly.

From Figure \ref{fig:machv}
we note that the average convective velocity increases rapidly within the first $2 \times 10^3 \s$.
The value then decreases and increases back,
followed by a decrease in the time interval $8 \times 10^3 -1.8 \times 10^4 \s$,
after which it increases until we terminate the simulation.
Variations over a timescale of $\approx 0.5-1 \times 10^4 \s$
are seen also in the values of the angular momentum in thick shells as presented in 
Figure \ref{fig:jfluctutations}.
Both figures \ref{fig:machv} and \ref{fig:jfluctutations}
seem to suggest that the simulation has reached a more or less steady state,
but with significant fluctuations.
It would be useful to extend the calculations to check that this is actually the case,
although this is computationally expensive.
The large amplitudes and the long timescale of the fluctuations
are the source of the relatively large value of the
stochastic specific angular momentum of the accreted gas.

Another indication that we have reached a more or less steady state is presented in Figure \ref{fig:jspecific}.
We see that the specific angular momentum components in thin shells reach maximum values
that are about equal to those of the standard deviation of specific angular momentum
calculated analytically from the MLT as in \cite{GilkisSoker2014}.
This estimation depends on the shell width, $\Delta r$,
which in the present work is larger than in \cite{GilkisSoker2014}.
In our previous work, the thin shells had a width equivalent to a difference of free-fall times
that is a small multiple of the Keplerian time for a NS,
which explains the higher values obtained for the standard deviation of specific angular momentum there.
Additionally, the analytic estimation using MLT values for the present work is somewhat problematic,
as the flow velocity is higher in the numerical simulation than in MLT.
On the other hand, the length scale taken may need to be smaller than the mixing-length,
which is close to the scale of vortices here, while the relevant length might be only part of a vortex.
These factors compensate each other to some extent,
so that the analytic estimation is close to the fluctuations of specific angular momentum in the simulation.
This argument is an order-of-magnitude one,
but nonetheless supports the claim that a steady state has been achieved.
The diameter of the vortex, for example, is taken to be of the order of the mixing-length
(or the radius of the vortex $a_c$ about equal to half the mixing-length).
This order of magnitude estimate, as presented in Figure \ref{fig:jspecific},
further demonstrates the merit of the analytic model.

\subsection{Convective luminosity}
\label{subsec:convlum}
 
To further reveal the convective properties of the 3D simulations
we study now the energy transport property of the convective motion.
We start by writing the three equations assumed by the MLT.
In doing so we follow the nomenclature of \cite{Kippenhahn2012}.
The convective flux is given by the heat capacity per unit mass of the gas $c_p$
multiplied by the mass flux $\rho v_c$,
where $v_c$ is the convective velocity according to the MLT,
and by the temperature difference $DT$
between the starting and end point of the convective element
\begin{equation}
F_\mathrm{conv} =  \rho v_\mathrm{c} c_p DT .
\label{eq:Conv1}
\end{equation}
The temperature difference is given by
\begin{equation}
DT= \left( \nabla- \nabla_e \right) \frac{l_m}{2} \frac{1}{H_p} T
\label{eq:Conv2}
\end{equation}
where $\nabla \equiv \partial \ln T / \partial \ln P$ is the
temperature gradient (with respect to pressure) in the star,
while $\nabla_e$ is the temperature gradient of the convective element;
$l_m$ is the mixing length and $H_p$ is the pressure scale height.
The velocity of the convective element is given by
\begin{equation}
v_\mathrm{c}^2  = g \delta \left( \nabla- \nabla_e \right) \frac{l_m^2}{8 H_p},
\label{eq:Conv3}
\end{equation}
where $\delta \equiv - \left(\partial \ln \rho / \partial \ln T\right)_{P}$;
for an ideal gas $\delta=1$.
 
We now manipulate these three equations as follows.
From equations (\ref{eq:Conv2}) and (\ref{eq:Conv3}) we derive
\begin{equation}
DT= \frac{4T}{g \delta l_m} v_\mathrm{c}^2
\label{eq:Conv4},
\end{equation}
which together with equation (\ref{eq:Conv1}) give
\begin{equation}
F_\mathrm{conv} =  \rho v_\mathrm{c}^3  \frac{4T c_P}{g \delta l_m} .
\label{eq:Conv5}
\end{equation}
The convective luminosity,
equals to the integration of the convective flux over a spherical surface,  
can be written then in the form
\begin{equation}
L_\mathrm{conv}=  \Lambda(r) L_\mathrm{KE} ,
\label{eq:Conv6}
\end{equation}
where $\Lambda\left(r\right) \equiv {4 T c_P}/{g \delta l_m}$
depends only on the radius.
In the MLT the kinetic luminosity $L_\mathrm{KE}$ is given by
\begin{equation}
L_\mathrm{KE-MLT} \left(r\right) \equiv 4 \pi r^2 \rho v_\mathrm{c}^3 .
\label{eq:Conv7}
\end{equation}
 
Let us summarize the important assumptions that went into the
derivation of equation (\ref{eq:Conv7}).
($i$) The convective elements perform radial motion.
($ii$) They are destroyed at a radius that is larger by $l_m$ from where they were formed.
($iii$) They transfer all their extra thermal energy where they are destroyed.
 
In the full 3D simulations we find that the motion is composed of vortices,
rather than radial motions (Fig. \ref{fig:velocitycloseup}).
This implies that the hotter gas that is flowing at the top of a vortex
loop is not destroyed, hence does not transfer to the surroundings
all of its thermal energy.
However, the net flux should be about equal to that in the MLT,
as the structure of the star is the same, hence same luminosity.
To compare the convective luminosity in our 3D simulation to that of the MLT,
we need to replace $L_\mathrm{KE-MLT}$ as given in eqaution (\ref{eq:Conv7}) by
\begin{equation}
L_\mathrm{KE-3D} \left(r\right) = r^2 \int_0^\pi \sin \theta d {\theta} \int_0^{2 \pi} d{\phi} \rho v^2 \left( \vec{v} \cdot \hat{r} \right). 
\label{eq:Conv8}
\end{equation}
where the density and velocity now depend on the location,
$\rho \left(r, \theta, \phi\right)$ and $\vec{v} \left(r, \theta, \phi\right)$,
respectively.
To take into account the numerical grid,
in calculating  $ L_\mathrm{KE-3D} \left(r\right)$ equation (\ref{eq:Conv8}) is replaced by
\begin{multline}
L_\mathrm{KE-3D} \left(r\right) =
\frac{1}{\Delta r} r^2 \\
\times \int_r^{r+\Delta r} dr^\prime \int_0^\pi \sin \theta d {\theta} \int_0^{2 \pi} d{\phi} \rho v^2  \left( \vec{v} \cdot \hat{r}^\prime \right). 
\label{eq:Conv9}
\end{multline}
 
\begin{figure}
   \centering
    \includegraphics*[scale=0.54]{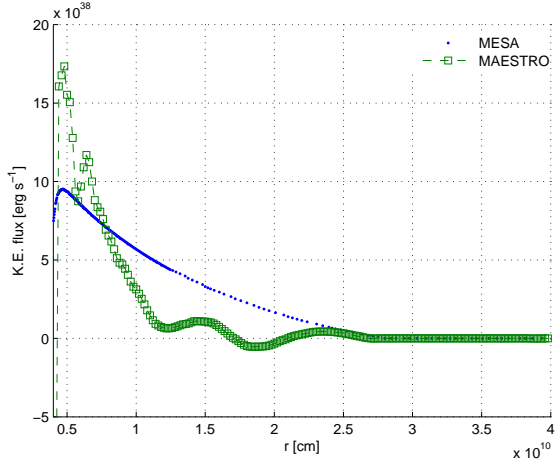} \\
\caption{Comparing the kinetic luminosity of the MLT according to
equation (\ref{eq:Conv7}), in blue dots with that of the full
3D simulation according to equation (\ref{eq:Conv9}), in green
squares. }
      \label{fig:Lconv}
\end{figure}

In Figure \ref{fig:Lconv} we plot the kinetic luminosity of the MLT
according to equation (\ref{eq:Conv7}), in blue dots, and that
of the full 3D simulation according to equation (\ref{eq:Conv9}),
in green squares.
Although the kinetic luminosity of the 3D simulation is not identical to that of the MLT,
it has the same bulk behavior.  
Our results at other times show a kinetic luminosity profile that changes with time.
For example, the peak near the inner boundary changes slightly,
such that the average value of the 3D results is less than twice that of the MLT near the peak of the kinetic luminosity.
Overall, Figure \ref{fig:Lconv} makes us confident in the convective properties revealed by our 3D simulations.
Most important,
the convective region is composed of large vortices that may translate after collapse to local fluctuations
in the specific angular momentum of the mass accreted by the newly born NS or BH.

Despite our finding that the typical convective velocity in the 3D simulation is much larger than the typical velocity in the MLT,
the convective luminosity of the two cases are very similar.
The reason is as follows.
In the MLT the convective velocity is calculated from the assumption
that hot convective cells deposit all their extra energy after they move a radial distance equal to the mixing-length.
This implies that the calculation of the convective luminosity of the MLT involves only upward moving gas.
In the full 3D simulations convective cells are not destroyed when they reach maximum radius in the convective zone.
They rather perform motions along vortices,
and deposit a small fraction of their extra energy when they reach the maximum radius in their cyclical motion.
Therefore,
the integral performed in equation (\ref{eq:Conv8}) (or eq. \ref{eq:Conv9}) involves a large flux downward.
Namely,
the factor $\vec{v} \cdot \hat{r} $ in those equations posses both positive
(for upward motion)
and negative
(for downward motion)
values.
The very large flux upward minus the very large flux downward gives a value of the convective luminosity about equal to that in the MLT,
as expected if the numerical simulation converges to a solution that yields the correct luminosity through the convective zone.

\section{IMPLICATIONS FOR CORE-COLLAPSE}
\label{sec:collapse}

The flow structure of the helium presented in section \ref{sec:results}
is probably similar to that just prior to collapse,
and we can look at possible implications  on the post-collapse dynamics.
The scale of the vortices is such that the difference between the free-fall times
onto the newly formed NS or black hole (BH), between the edges of a vortex, is about tens of seconds.
The local specific angular momentum as shown in Figure \ref{fig:velocitymap}
is higher than the Keplerian specific angular momentum for orbit around a newly-formed NS.
The large vortices imply then that on a short time scale of $\approx 0.01 \s$,
equal to several dynamical times on the newly born NS or BH,
a small number of vortices contribute to the accretion,
and only a small fraction from a vortex is accreted,
contributing a small mass with high specific angular momentum.
The differences in free-fall times between shells with differing angular momentum axes,
equivalent to many orbital times close to compact object (where we assume jets are launched),
allow for sufficient time for formation of disks and jets.
The average value obtained for shells as described in Figure \ref{fig:jspecific} is lower,
and may lead to the formation of
a thick accretion disk (or an accretion belt;
for further discussion of this general effect see \citealt{GilkisSoker2015} and \citealt{PapishGilkisSoker2015}).

Considering jittering jets, which we suggest explode the star,
we can make the following crude energy estimate.
We take only shells with the high specific angular momentum to form accretion belts and launch jets.
Say we take only parts with $j > 6 \times 10^{14} \cm^2 \s^{-1}$ according to Figure \ref{fig:jspecific};
the fluctuations in narrower shells are larger even.
The mass in these parts is $\approx 0.1 M_\odot$.
Accretion of this mass onto the NS of a mass of $\approx 2 M_\odot$, or later a BH,
can be channeled into jets with energy $\approx 10 \%$ of the rest mass of the shell.
This amounts to $\approx 2 \times 10^{52} \erg$.
To account for a CCSNe it is sufficient that only about 10 percent of the time the accretion belt/disk exists and launches jets 
\citep{PapishSoker2011}.
We emphasize that this last calculation is approximate and somewhat speculative.
However, it is worth consideration as our main motivation for studying angular momentum fluctuations
in the helium shell was implications for jet-driven explosion mechanisms.
Due to the high resolution required,
the need to explore convective regions closer to the iron-core,
and the uncertainties of the jet-formation mechanism,
we cannot yet reach definitive conclusions regarding the implications of convective regions
for the jittering jets model.

\section{DISCUSSION AND SUMMARY}
\label{sec:summary}

We have presented a novel study of angular momentum fluctuations in stellar convective regions,
using the sub-sonic solver MAESTRO.
Further refinement and development of this method may enable the study of such fluctuations in different
stellar models at different stages (e.g., just prior to core-collapse).
These fluctuations may have
implications on the core-collapse explosion mechanism,
but may also be relevant for
the process of stochastic spin-up of the core \citep{Fuller2015}.
In the present study we found large-scale fluctuations of angular momentum in the convective helium zone of an evolved massive star,
while the energy transport is similar to that expected from MLT, strengthening the validity of our simulation.
Future studies will explore similar phenomena in silicon and oxygen shells (as done by, e.g., \citealt{Mueller2016}),
and effects of rotation (as done by, e.g., \citealt{Chatzopoulos2016}).

In this study we focused on the convective helium zone surrounding an oxygen core.
During the remainder of the stellar evolution time,
the late burning stages eventually produce an iron core surrounded by silicon and oxygen shells.
We assume that the structure of the helium shell does not change much in the relatively short time left to collapse.
This implies that our results are relevant for the final stage of the star,
although there may be late-stage processes beyond our one-dimensional stellar modeling,
such as significant wave-driven mass-loss \citep{Quataert2012,Shiode2014} or expansion \citep{Mcley2014}.

The mass inner to the convective region studied here is $M_\mathrm{in} \approx 2.4 M_\odot$,
and a jet-driven explosion powered by accretion from the helium zone will leave a heavy NS.
Loss of mass-energy to escaping neutrinos may reduce the remnant mass to around $M_\mathrm{NS} \approx 2 M_\odot$.
For a lower mass NS to form (e.g., with $M_\mathrm{NS} \approx 1.4 M_\odot$),
a jet-driven explosion will have to be powered by accretion from the silicon/oxygen shells.
For more massive stars,
the helium shell may surround a core massive enough to form a BH upon collapse.
If the accretion of helium powers an explosion,
then the relevance of
`failed supernovae' \citep{Kochanek2008,OConnor2011,Kochanek2014},
or `very weak supernovae' \citep{Lovegrove2013}
may be lessened.
A supernova leaving a BH remnant may be prevalent for very massive stars --
this may be the case for, e.g., SN 2005gl \citep{GalYam2009}.

We can consider the present results on the source of   stochastic accreted angular momentum  from the helium  shell
in combination with our previous results on a possible stochastic angular momentum accretion from the silicon shell
\citep{GilkisSoker2015},
and the results of \cite{PapishGilkisSoker2015}
regarding the stochastic angular momentum accretion as a result of the standing accretion-shock instability (SASI).
We then come at an emerging idea where the newly formed NS or BH accretes mass with notable
specific angular momentum fluctuations.
These accretion episodes might lead to the formation of intermittent accretion disks or accretion belts
that launch jets \citep{Schreier2016}.
The present study is another step in developing the jittering jets model for CCSNe.
This is a preliminary study, and the presented results will have to be scrutinized with higher-resolution simulations.
Also, in future works we will need to consider the combined effect of the different sources of stochastic angular momentum,
as well as the angular momentum contribution from rotation.
Magnetic fields may also be relevant for the pre-collapse evolution of stochastic angular momentum,
and they are most definitely important for the post-collapse possibility of jet formation.

\section*{Acknowledgments}

We thank an anonymous referee for helpful comments that improved the presentation of our results.
We thank Michael Zingale for helpful correspondence regarding the MAESTRO code,
as well as the rest of the developers of MAESTRO for making it accessible and publicly available for the scientific community.
Simulations were ran on the Israeli astrophysics I-CORE astric HPC.

\appendix
\section{Lower resolution simulations}
\label{app:low}

To check the effects of the simulation resolution on the angular momentum fluctuations,
we performed additional simulations
with the best resolution of the  AMR grid (termed finest-resolution)
lower than our nominal simulation.
Our computational resources prohibit us from increasing the resolution in a full simulation
needed when looking at angular momentum of entire shells.
Other properties of the flow are studied in higher resolution octant simulations.
This is presented in Appendix \ref{app:oct}.

Figure \ref{fig:jabs} shows the magnitude of the specific angular momentum in different shells
for simulations of varying finest-resolutions.
The behavior and magnitude are similar, although hard to quantify, due to the stochastic nature of the angular momentum in the shells.
Figure \ref{fig:fullres} show the resolution dependency
of the time-averaged value of the specific angular momentum in different shells,
and of the time-averaged maximal Mach number.
As in Figure \ref{fig:jabs}, the effect of resolution on the specific angular momentum is not entirely clear.
The maximal Mach number, however, varies very regularly with the changing resolution.
Extrapolating linearly to infinite resolution reduces the Mach number by a factor of $\approx 2$
compared to our nominal simulation.
We might expect the angular momentum to change accordingly,
but this must be studied in simulations of higher resolution,
beyond our current computational resources.
\begin{figure}
   \centering
    \includegraphics*[scale=0.34]{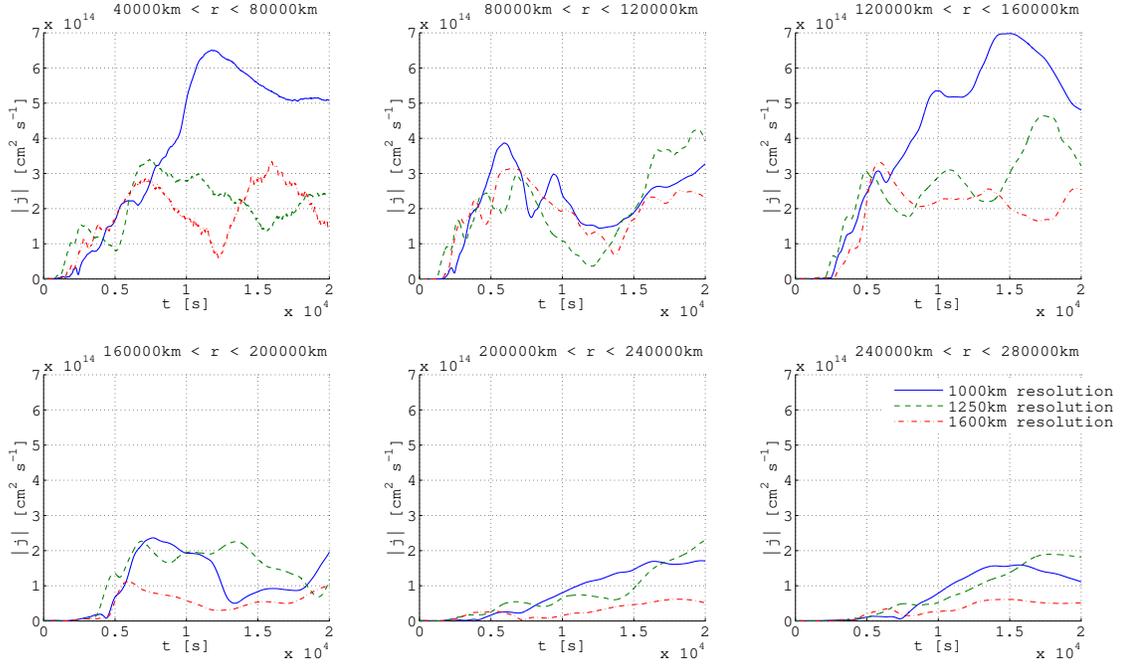} \\
      \caption{Specific angular momentum magnitude for different shells.
      The radial extent of the shells is detailed above each panel.
      Results from three different full-space simulations are shown,
      with finest-resolutions of $1000\km$
      (the finest-resolution in our nominal simulation, analyzed in the main body of the paper),
      $1250\km$ and $1600\km$.}
      \label{fig:jabs}
\end{figure}
\ifmnras
\begin{figure*}
\else
\begin{figure}
\fi
\begin{tabular}{cc}
\ifmnras
{\includegraphics*[scale=0.53]{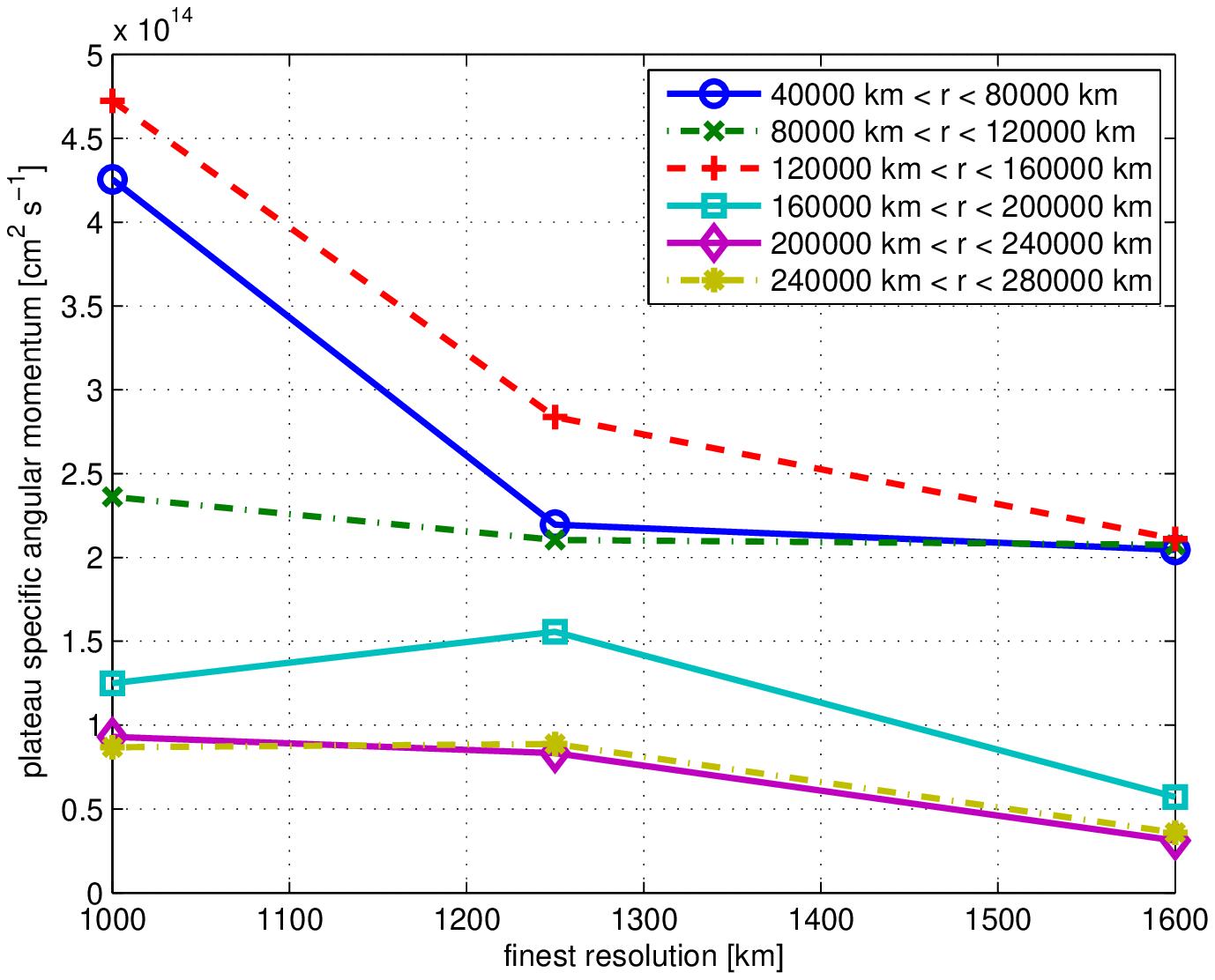}} &
{\includegraphics*[scale=0.53]{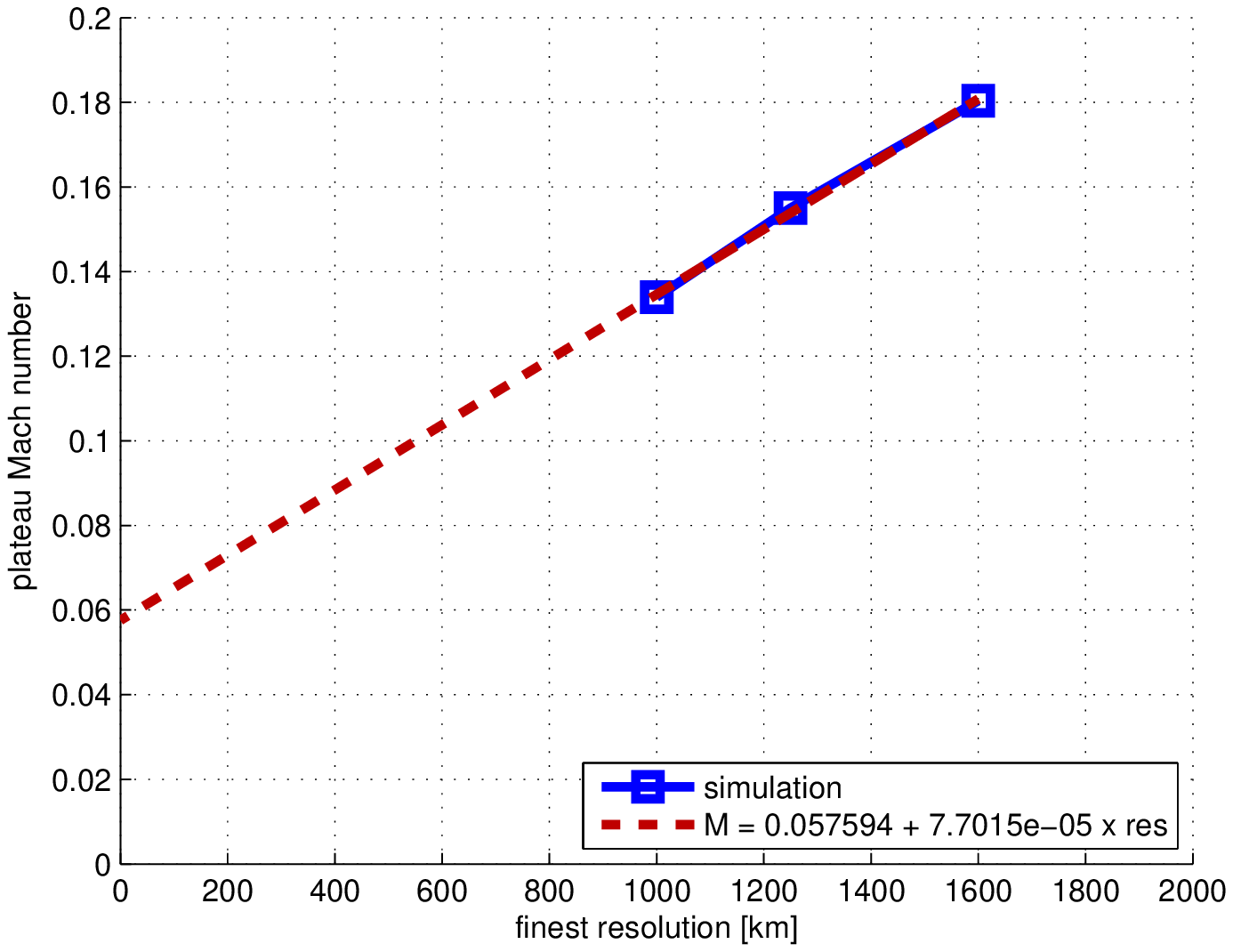}} \\
\else
{\includegraphics*[scale=0.54]{jplat_res.eps}} &
{\includegraphics*[scale=0.54]{mach_res_full.eps}} \\
\fi
\end{tabular}
      \caption{\textit{Left:} Average over time (for $t>3000\s$)
      of the specific angular momentum in different shells for various finest-resolutions in full-space simulations.
      \textit{Right:} Average value of the maximal Mach number for $t>3000\s$
      as a function of finest grid resolution.
      The presented linear fit shows the extrapolated value of the Mach number, $\mathcal{M} \rightarrow 0.058$.}
      \label{fig:fullres}
\ifmnras
\end{figure*}
\else
\end{figure}
\fi

\section{Octant simulations}
\label{app:oct}

\begin{figure}
   \centering
    \includegraphics*[scale=0.42]{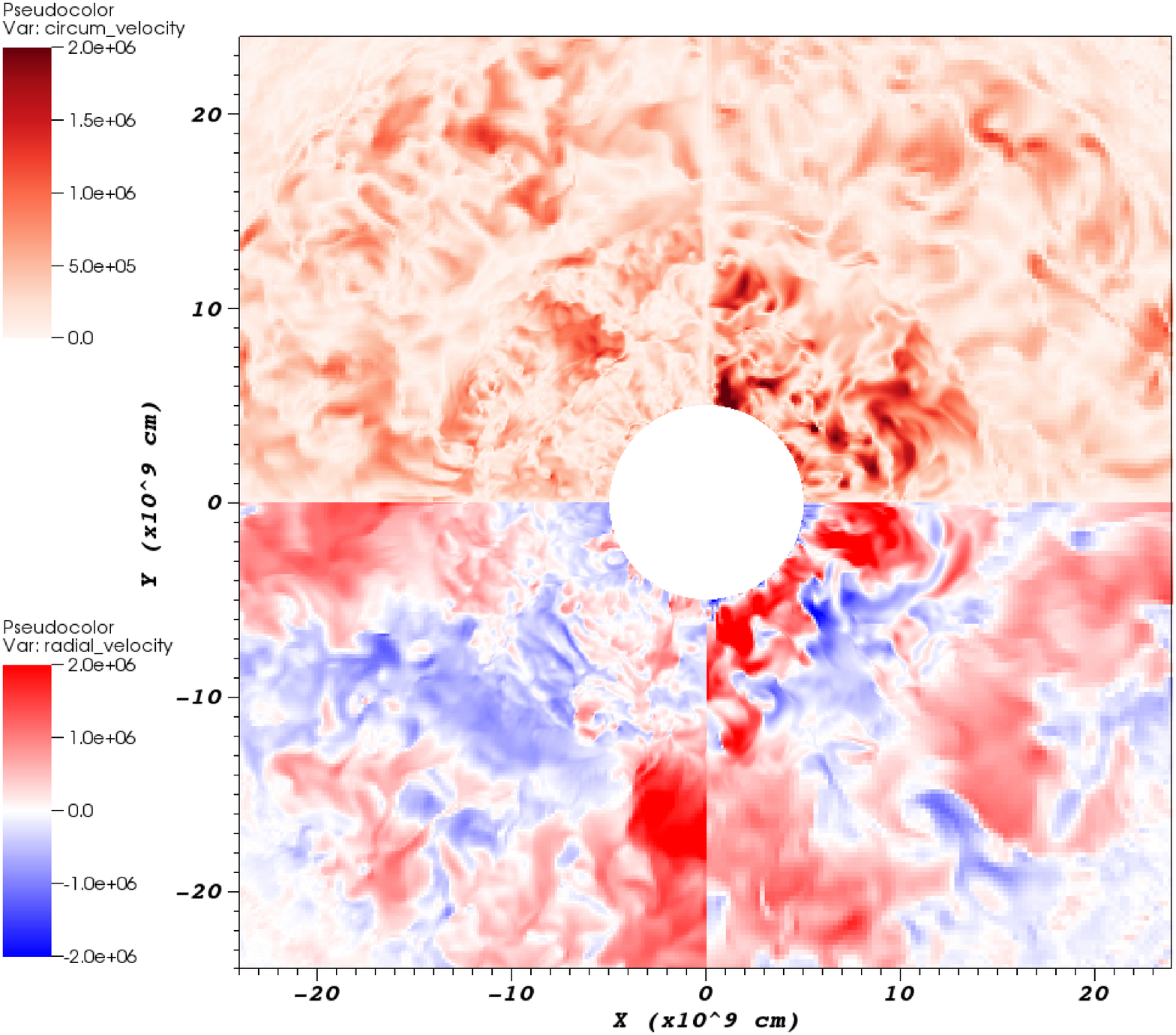} \\
      \caption{Presentation of two variables in two simulations in four different quarters.
      The two simulations are performed in one octant of the full space due to limited computer resources.
      All velocities are in units of $\cms$ (so that the highest velocity magnitude presented is $2\times 10^6 \cms$).
      \textit{Top:} Circumferential velocity (i.e., perpendicular to the radial direction)
      with finest-resolution of $800\km$ (left) and $1250\km$ (right).
      Light shades of orange represent lower velocities, and darker shades represent high circumferential flow velocities.
      \textit{Bottom:} Radial velocity with finest-resolution of $800\km$
      (left) and $1250\km$ (right). Red shades represent outward motion, and blue shades represent inward motion.
      The inner oxygen core is omitted from the figure.}
      \label{fig:vel4octs}
\end{figure}

To further investigate the issue of convergence,
we performed several simulations of one eighth of the full simulation with reflecting boundary conditions (`octant' simulations),
with different maximal simulation resolutions.
Two simulations, with finest resolutions of $800\km$ and $1250\km$,
are presented at $t=20000\s$ in Figure \ref{fig:vel4octs}.
From Figure \ref{fig:vel4octs} we can see that the outer part of the convective region is similar in the two simulations,
while in the inner part the simulation with lower finest-resolution the flow velocity is somewhat higher.
We also see by comparing the upper to lower panel on each side
that the magnitude of the radial velocity component and the magnitude of the circumferential velocity component are similar.
This shows that the convective flow is well-developed.

Figure \ref{fig:ekradcircum} shows the 
kinetic energy due to radial component of the velocity only and due to the circumferential component of the velocity only
in the convective region,
for octant simulations with three different finest-resolutions, as well as our nominal full simulation, at $t=20000\s$.
The kinetic energy in the outer part seems converged,
while the inner part is sensitive to the simulation resolution.
The radial and circumferential flow components of the kinetic energy are nearly equal,
correspondingly with the velocity flows seen in Figure \ref{fig:vel4octs}.
\ifmnras
\begin{figure*}
\else
\begin{figure}
\fi
\begin{tabular}{cc}
\ifmnras
{\includegraphics*[scale=0.53]{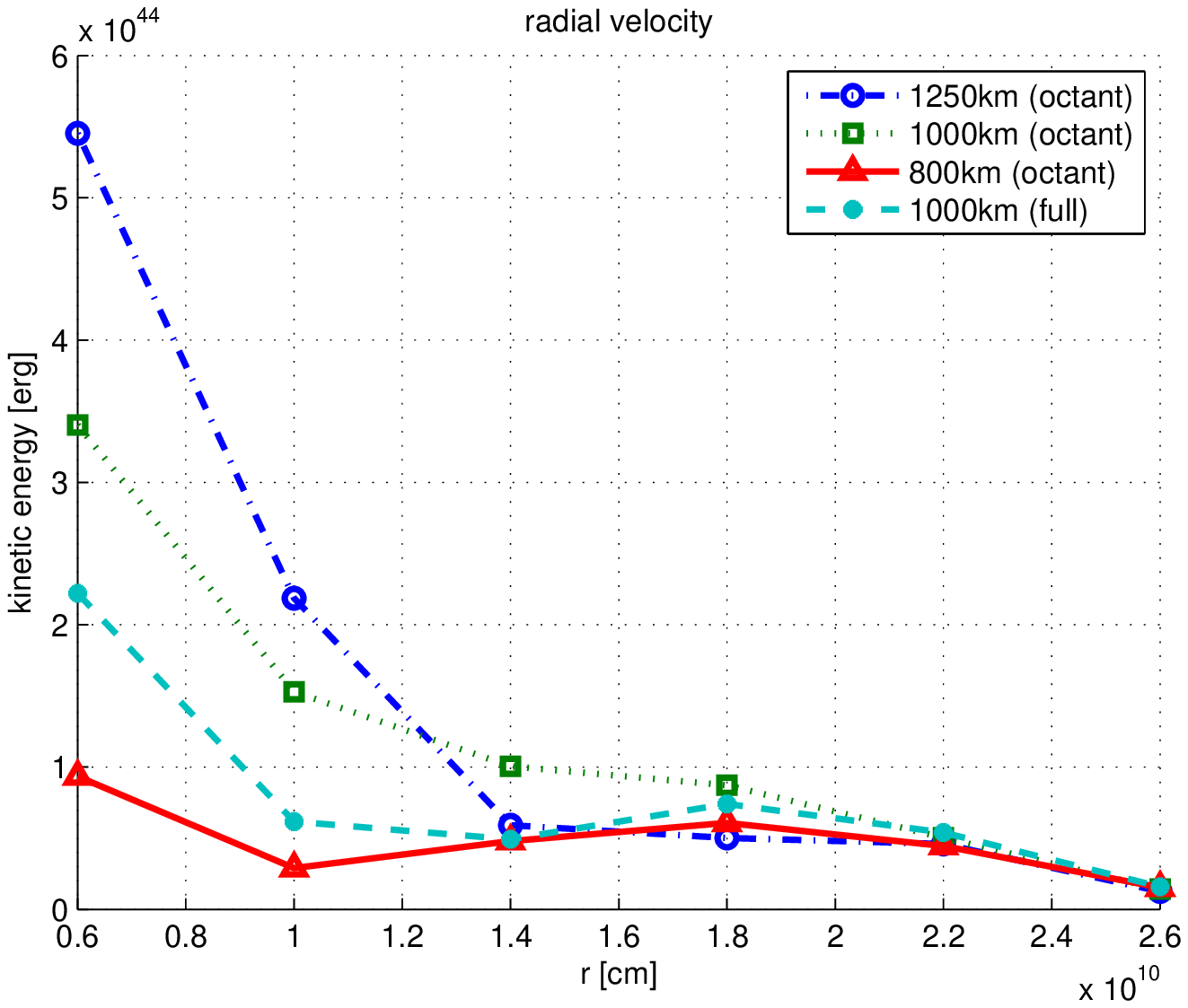}} &
{\includegraphics*[scale=0.53]{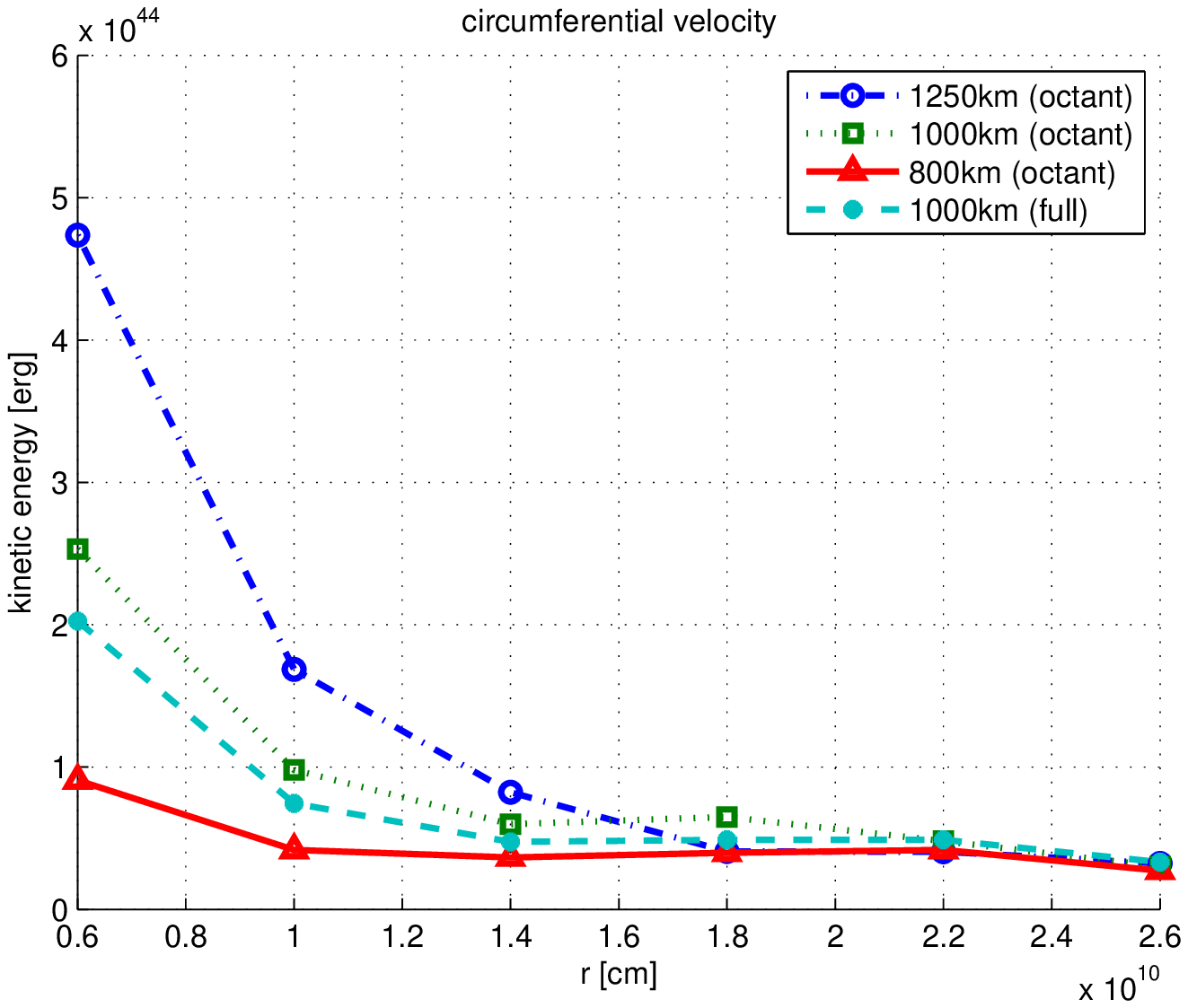}} \\
\else
{\includegraphics*[scale=0.54]{EkRad.eps}} &
{\includegraphics*[scale=0.54]{EkCircum.eps}} \\
\fi
\end{tabular}
      \caption{\textit{Left:} Kinetic energy of the radial velocity component for six different shells
      in octant simulations of varying finest resolution.
      The values for finest resolutions of $800\km$ and $1250\km$
      are obtained for the simulations shown in Figure \ref{fig:vel4octs}.
      The kinetic energy is multiplied by eight for the presentation of the octant simulations.
      \textit{Right:} Kinetic energy of the circumferential velocity component (i.e., perpendicular to the radial direction)
      for six different shells
      in octant simulations of varying finest resolution.
      The $r$ coordinate is taken in the middle of each shell.      
      The masses of the shells from the inner shell to the outer are
      $0.25$, $0.25$, $0.23$, $0.22$, $0.19$ and $0.17 M_\odot$.}
      \label{fig:ekradcircum}
\ifmnras
\end{figure*}
\else
\end{figure}
\fi

Figure \ref{fig:octres} shows the kinetic energy in the region between
$r=40000\km$ and $r=280000\km$ as a function of simulation time.
The kinetic energy in the simulations has a transient phase with a large value,
and then goes down and plateaus.
This behavior is similar to that obtained by \cite{Viallet2013} in their simulations of a convective envelope of a red giant star
(see their Figure 2).
It can be seen that the simulations are not converged,
as the kinetic energy plateau is lower for more highly-resolved simulations, as is the time-averaged maximal Mach number.
Extrapolating from the values of the maximal Mach number to infinite resolution reduces the Mach number by a factor of 2--3 compared to
our nominal simulation, similar to the results of the full simulations with lower resolution discussed in Appendix \ref{app:low}.
\ifmnras
\begin{figure*}
\else
\begin{figure}
\fi
\begin{tabular}{cc}
\ifmnras
{\includegraphics*[scale=0.53]{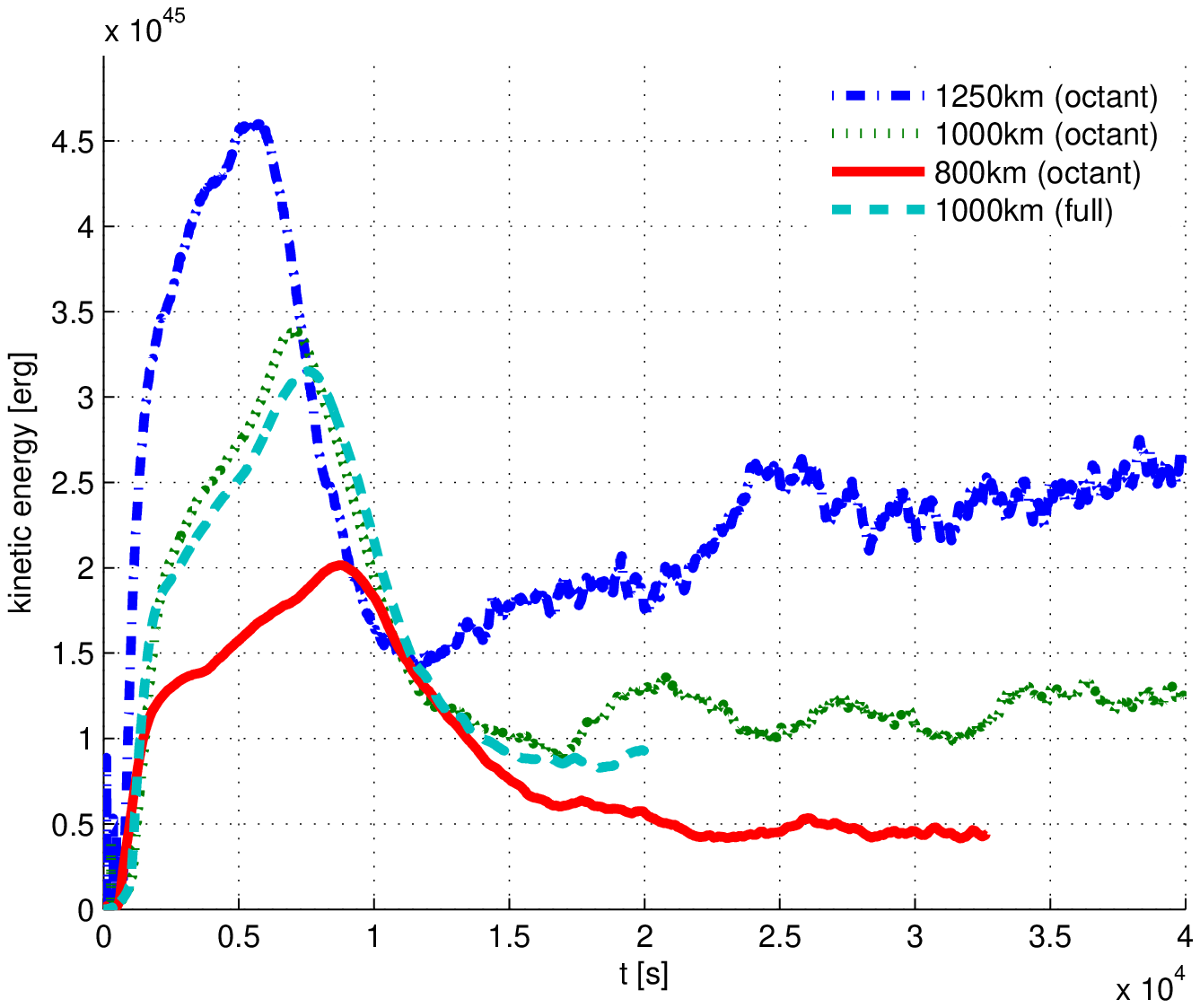}} &
{\includegraphics*[scale=0.53]{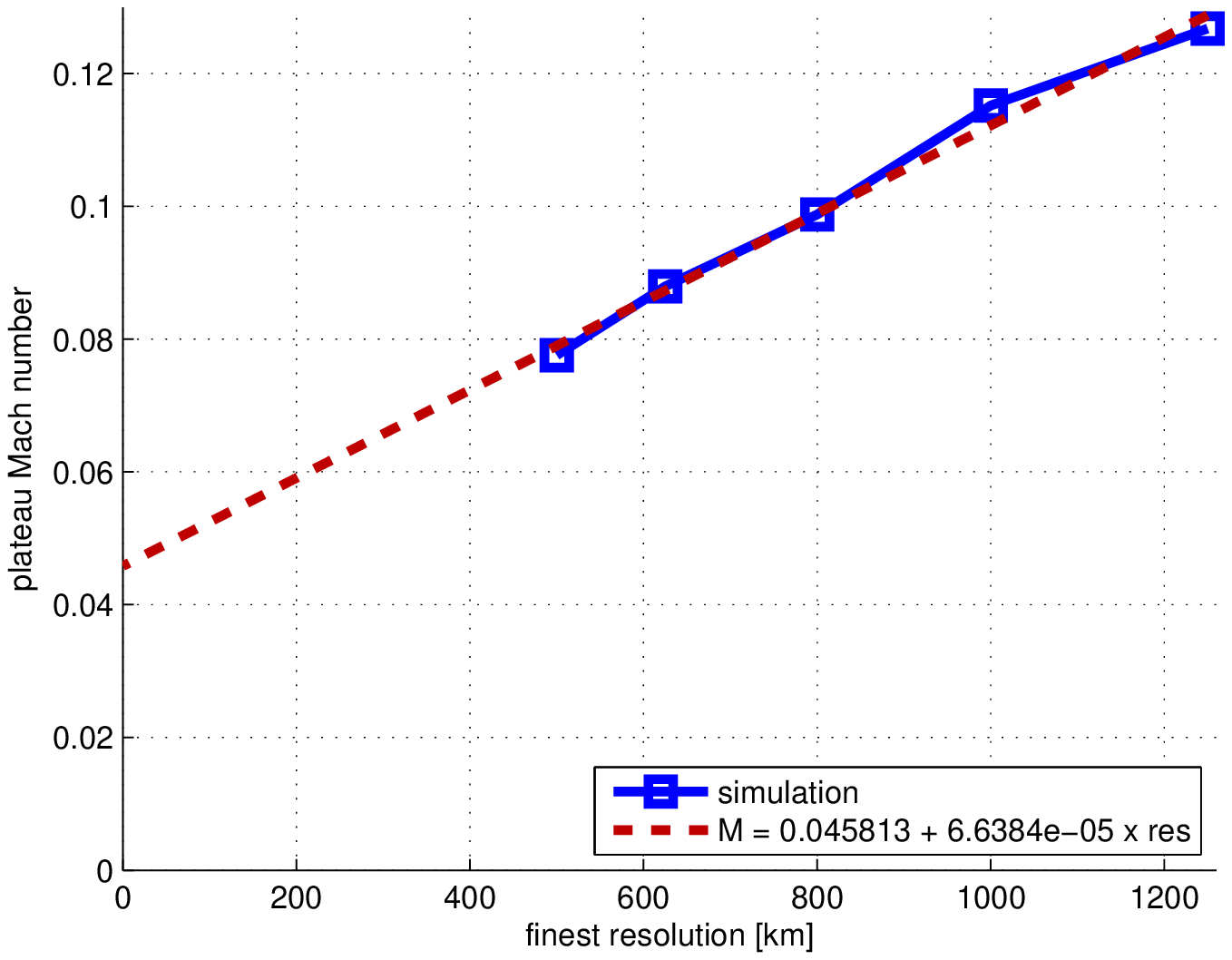}} \\
\else
{\includegraphics*[scale=0.54]{ek_res_full.eps}} &
{\includegraphics*[scale=0.54]{mach_res.eps}} \\
\fi
\end{tabular}
      \caption{\textit{Left:} Kinetic energy of the convective 
      flow as function of time for our
      nominal simulation,
      and for three octant simulations with different resolutions (the kinetic energy is multiplied by eight for the presentation of the octant simulations).
      \textit{Right:} Average value of the maximal Mach number for $t>3000\s$
      as a function of finest grid resolution.
      The presented linear fit shows the extrapolated value of the Mach number, $\mathcal{M} \rightarrow 0.046$.}
      \label{fig:octres}
\ifmnras
\end{figure*}
\else
\end{figure}
\fi

\ifmnras
	\bibliographystyle{mn2e}

\begin{thebibliography}{99}
\else
	\begin{thebibliography}{}\addcontentsline{toc}{section}{References}
\fi


\bibitem[Almgren et al.(2006a)]{Almgren2006a} Almgren, A.~S., Bell, J.~B., Rendleman, C.~A., \& Zingale, M.\ 2006a, \apj, 637, 922

\bibitem[Almgren et al.(2006b)]{Almgren2006b} Almgren, A.~S., Bell, J.~B., Rendleman, C.~A., \& Zingale, M.\ 2006b, \apj, 649, 927

\bibitem[Almgren et al.(2008)]{Almgren2008} Almgren, A.~S., Bell, J.~B., Nonaka, A., \& Zingale, M.\ 2008, \apj, 684, 449

\bibitem[Akiyama et al.(2003)]{Akiyama2003} Akiyama, S., Wheeler, J.~C., Meier, D.~L., Lichtenstadt, I.\ 2003, \apj, 584, 954


\bibitem[Arnett et. al(2015)]{Arnett2015} {Arnett}, D., {Meakin}, C., Viallet, M., Campbell, S.~W., Lattanzio, J.~C., \& {Moc{\'a}k}, M. \ 2015, \apj, 809, 30

\bibitem[Arnett et. al(2009)]{Arnett2009} {Arnett}, D., {Meakin}, C., \& {Young}, P.~A. \ 2009, \apj, 690, 1715



\bibitem[Beeck et al.(2015)]{Beeck2015} Beeck, B., {Sch{\"u}ssler}, M., Cameron, R.~H., \& Reiners, A.\ 2015, \aap, 581, 42

\bibitem[Bethe \& Wilson(1985)]{bethe1985} Bethe, H.~A., \& Wilson, J.~R.\ 1985, \apj, 295, 14

\bibitem[Blondin \& Mezzacappa(2007)]{BlondinMezzacappa2007} Blondin, J.~M., \& Mezzacappa, A.\ 2007, \nat, 445, 58

\bibitem[Blondin, Mezzacappa \& DeMarino(2003)]{BlondinMezzacappa2003} Blondin, J.~M., Mezzacappa, A., \& DeMarino, C.\ 2003, \apj, 584, 971







\bibitem[Burrows, Hayes \& Fryxell(1995)]{Burrows1995} Burrows, A., Hayes, J., \& Fryxell, B.~A.\ 1995, \apj, 450, 830


\bibitem[Chatzopoulos et al.(2016)]{Chatzopoulos2016} Chatzopoulos, E., Couch, S.~M., Arnett, W.~D., Timmes, F.~X.\ 2016, \apj, 822, 61

\bibitem[Colgate \& White(1966)]{Colgate1966} Colgate, S.~A., \& White, R.~H.\ 1966, \apj, 143, 626


\bibitem[Couch et al.(2015)]{Couchetal2015} Couch, S.~M., Chatzopoulos, E., Arnett, W.~D., \& Timmes, F.~X.\ 2015, \apjl, 808, L21


\bibitem[Couch \& Ott(2013)]{CouchOtt2013} Couch, S.~M., \& Ott, C.~D.\ 2013, \apjl, 778, L7

\bibitem[Couch \& Ott(2015)]{CouchOtt2015} Couch, S.~M., \& Ott, C.~D.\ 2015, \apj, 799, 5

	

\bibitem[Fern{\'a}ndez(2010)]{Fernandez2010} Fern{\'a}ndez, R.\ 2010, \apj, 725, 1563

\bibitem[Fuller et al.(2015)]{Fuller2015} Fuller, J., Cantiello, M., Lecoanet, D., \& Quataert, E.\ 2015, \apj, 810, 101


\bibitem[Gal-Yam \& Leonard(2009)]{GalYam2009} {Gal-Yam}, A., \& {Leonard}, D.~C.\ 2009, \nat, 458, 865

\bibitem[Gilet et al.(2013)]{Gilet2013} Gilet, C., Almgren, A.~S., Bell, J.~B., Nonaka, A., Woosley, S.~E., \& Zingale, M.\ 2013, \apj, 773, 137
  
\bibitem[Gilkis \& Soker(2014)]{GilkisSoker2014}  Gilkis, A. \& Soker, N.\ 2014, \mnras, 439, 4011

\bibitem[Gilkis \& Soker(2015)]{GilkisSoker2015}  Gilkis, A. \& Soker, N.\ 2015, \apj, 806, 28



\bibitem[Janka(2012)]{Janka2012} Janka, H.-T.\ 2012, Annual Review of Nuclear and Particle Science, 62, 407


\bibitem[Khokhlov et al.(1999)]{Khokhlov1999} Khokhlov, A.~M., H{\"o}flich, P.~A., Oran, E.~S., et al.\ 1999, \apjl, 524, L107

\bibitem[Kippenhahn et al.(2012)Kippenhahn, Weigert \& Weis]{Kippenhahn2012} Kippenhahn, R., Weigert, A., \& Weiss, A.\ 2012, Stellar Structure and Evolution, Springer-Verlag Berlin Heidelberg


\bibitem[Kochanek(2014)]{Kochanek2014} Kochanek, C.~S.\ 2014, \apj, 785, 28

\bibitem[Kochanek et al.(2008)]{Kochanek2008} Kochanek, C.~S., Beacom, J.~F., Kistler, M.~D., Prieto, J.~L., Stanek, K.~Z., Thompson, T.~A., \& Y{\"u}ksel, H.\ 2008, \apj, 684, 1336


\bibitem[Kushnir(2015a)]{Kushnir2015a} Kushnir, D.\ 2015a, \href{http://arxiv.org/abs/1502.03111}{arXiv:1502.03111}

\bibitem[Kushnir(2015b)]{Kushnir2015b} Kushnir, D.\ 2015b, \href{http://arxiv.org/abs/1502.03111}{arXiv:1506.02655}

\bibitem[Kushnir \& Katz(2015)]{KushnirKatz2014} Kushnir, D., \& Katz, B.\ 2015, \apj, 811, 97

\bibitem[Lazzati et al.(2012)]{Lazzati2012} Lazzati, D., Morsony, B.~J., Blackwell, C.~H., \& Begelman, M.~C.\ 2012, \apj, 750, 68

\bibitem[LeBlanc \& Wilson(1970)]{LeBlanc1970} LeBlanc, J.~M., \& Wilson, J.~R.\ 1970, \apj, 161, 541

\bibitem[Liebend{\"o}rfer et al.(2005)]{Liebend2005} Liebend{\"o}rfer, M., Rampp, M., Janka, H.-T., \& Mezzacappa, A.\ 2005, \apj, 620, 840

\bibitem[Lovegrove \& Woosley(2013)]{Lovegrove2013} Lovegrove, E., \& Woosley, S.~E.\ 2013, \apj, 769, 109

\bibitem[Malone et al.(2011)]{Malone2011} Malone, C.~M., Nonaka, A., Almgren, A.~S., Bell, J.~B., \& Zingale, M.\ 2011, \apj, 728, 118

\bibitem[Malone et al.(2014)]{Malone2014} Malone, C.~M., Zingale, M., Nonaka, A., Almgren, A.~S., \& Bell, J.~B.\ 2014, \apj, 788, 115
	


\bibitem[Mcley \& Soker(2014)]{Mcley2014} Mcley, L., \& Soker, N.\ 2014, \mnras, 445, 2492

\bibitem[Meakin \& Arnett(2007)]{Meakin2007} Meakin, C.~A., \& Arnett, D.\ 2007, \apj, 667, 448
   

\bibitem[Mezzacappa et al.(2001)]{Mezzacappa2001} Mezzacappa, A., Liebend{\"o}rfer, M., Messer, O.~E., et al.\ 2001, Physical Review Letters, 86, 1935


\bibitem[M{\"o}sta et al.(2015)]{Mosta2015} {M{\"o}sta}, P., Ott, C.~D., Radice, D., Roberts, L.~F., Schnetter, E., Haas, R.\ 2015, \nat, 528, 376


\bibitem[Mueller \& Janka(2015)]{MuellerJanka2015} Mueller, B., \& Janka, H.-T.\ 2015, \mnras, 448, 2141 


\bibitem[Mueller et al.(2016)]{Mueller2016} Mueller, B., Viallet, M., Heger, A., \& Janka, H.-T.\ 2016, \href{http://arxiv.org/abs/1605.01393}{arXiv:1605.01393}

\bibitem[Nadezhin(1980)]{Nadezhin1980} Nadezhin, D.~K.\ 1980, \apss, 69, 115


\bibitem[Nonaka et al.(2010)]{Nonaka2010} Nonaka, A., Almgren, A.~S., Bell, J.~B., Lijewski, M.~J., Malone, C.~M., \& Zingale, M.\ 2010, \apjs, 188, 358
  

\bibitem[Nugis \& Lamers(2000)]{Nugis2000} {Nugis}, T. \& {Lamers}, H.~J.~G.~L.~M.\ 2000, \aap, 360, 227

\bibitem[O'Connor \& Ott(2011)]{OConnor2011} {O'Connor}, E., \& Ott, C.~D.\ 2011, \apj, 730, 70


\bibitem[Papish, Gilkis \& Soker(2015)]{PapishGilkisSoker2015} Papish, O., Gilkis, A., \& Soker, N.\ 2015, \href{http://arxiv.org/abs/1508.00218}{arXiv:1508.00218}

\bibitem[Papish, Nordhaus \& Soker(2015)]{PapishNordhausSoker2015} Papish, O., Nordhaus, J., \& Soker, N.\ 2015, \mnras, 448, 2362

\bibitem[Papish \& Soker(2011)]{PapishSoker2011} Papish, O., \& Soker, N.\ 2011, \mnras, 416, 1697

\bibitem[Papish \& Soker(2012a)]{PapishSoker2012a} Papish, O., \& Soker, N.\ 2012a, in Roming P. W. A., Kawai N., Pian E., eds, Proc. IAU Symp. 279, Death of Massive Stars: Supernovae and Gamma-Ray Bursts. Cambridge Univ. Press, Cambridge, p. 377

\bibitem[Papish \& Soker(2012b)]{PapishSoker2012b} Papish, O., \& Soker, N.\ 2012b, \mnras, 421, 2763

\bibitem[Papish \& Soker(2014a)]{PapishSoker2014a} Papish, O., \& Soker, N.\ 2014a, \mnras, 438, 1027

\bibitem[Papish \& Soker(2014b)]{PapishSoker2014b} Papish, O., \& Soker, N.\ 2014b, \mnras, 443, 664

\bibitem[Paxton et al.(2011)]{Paxton2011} Paxton, B., Bildsten, L., Dotter, A., et al.\ 2011, \apjs, 192, 3

\bibitem[Paxton et al.(2013)]{Paxton2013} Paxton, B., Cantiello, M., Arras, P., et al. \ 2013, \apjs, 208, 4

\bibitem[Quataert \& Shiode(2012)]{Quataert2012} Quataert, E., \& Shiode, J. \ 2012, \mnras, 423, L92

\bibitem[Rampp \& Janka(2000)]{Rampp2000} Rampp, M., \& Janka, H.-T.\ 2000, \apjl, 539, L33


\bibitem[Schreier \& Soker(2016)]{Schreier2016} Schreier, R., \& Soker, N.\ 2016, \raa, 16e, 1

\bibitem[Shiode \& Quataert(2014)]{Shiode2014} Shiode, J.~H., \& Quataert, E.\ 2014, \apj, 780, 96

\bibitem[Soker(2010)]{Soker2010} Soker, N.\ 2010, \mnras, 401, 2793


\bibitem[Viallet et al.(2016)]{Viallet2016} Viallet, M., Goffrey, T., Baraffe, I., Folini, D., Geroux, C., Popov, M.~V., Pratt, J., \& Walder, R.\ 2016, \aap, 586, 153

\bibitem[Viallet et al.(2013)]{Viallet2013} Viallet, M., Meakin, C., Arnett, D. \& {Moc{\'a}k}, M.\ 2013, \apj, 769, 1

\bibitem[Viallet et al.(2015)]{Viallet2015} Viallet, M., Meakin, C., Prat, V. \& Arnett, D.\ 2015, \aap, 580, 61
   
\bibitem[Vink, de Koter \& Lamers(2001)]{Vink2001} {Vink}, J.~S., {de Koter}, A., \& {Lamers}, H.~J.~G.~L.~M.\ 2001, \aap, 369, 574

\bibitem[Zingale et al.(2009)]{Zingale2009} Zingale, M., Almgren, A.~S., Bell, J.~B., Nonaka, A., \& Woosley, S.~E.\ 2009, \apj, 704, 196

\bibitem[Zingale et al.(2015)]{Zingale2015} Zingale, M., Malone, C.~M., Nonaka, A., Almgren, A.~S., \& Bell, J.~B.\ 2015, \apj, 807, 60

\end{thebibliography}

\label{lastpage}

\end{document}